\documentclass[10pt,a4paper]{article}
\usepackage[utf8]{inputenc}
\usepackage{amsmath}
\usepackage{amsfonts}
\usepackage{amssymb}
\usepackage{xfrac}
\usepackage{bm}
\usepackage{upgreek}
\usepackage{graphicx}
\usepackage[all]{nowidow}
\usepackage[left=2.5cm,right=2.5cm,top=2.5cm,bottom=2.5cm]{geometry}
\usepackage{cite}
\usepackage{dirtytalk}
\usepackage{float}
\usepackage[doublespacing]{setspace}
\usepackage{authblk}
\usepackage{setspace}
\usepackage{color,soul}
\usepackage{xcolor}
\usepackage{pdfpages}
\usepackage{hhline}

\begin{document}

\title{Mixing Rule for Calculating the Effective Refractive Index Beyond the Limit of Small Particles}

\author[1]{Dominic T. Meiers} 
\author[1,2]{Georg von Freymann}

\affil[1]{Physics Department and Research Center OPTIMAS, RPTU Kaiserslautern-Landau, 67663 Kaiserslautern, Germany}
\affil[2]{Fraunhofer Institute for Industrial Mathematics ITWM, 67663 Kaiserslautern, Germany}

\maketitle
\newpage

\onehalfspacing

\begin{abstract} 
Considering light transport in disordered media, the medium is often treated as an effective medium requiring accurate evaluation of an effective refractive index. Because of its simplicity, the Maxwell-Garnett (MG) mixing rule is widely used, although its restriction to particles much smaller than the wavelength is rarely satisfied. Using 3D finite-difference time-domain simulations, we show that the MG theory indeed fails for large particles. Systematic investigation of size effects reveals that the effective refractive index can be instead approximated by a quadratic polynomial whose coefficients are given by an empirical formula. Hence, a simple mixing rule is derived which clearly outperforms established mixing rules for composite media containing large particles, a common condition in natural disordered media.

\end{abstract}

\section{Introduction}
Since perfect order is rarely found in natural systems, the interaction of electromagnetic waves with disordered materials is part of many sensing applications ranging from biophotonics \cite{Hernandez20,Spathmann15} over geophysics \cite{Cosenza09,Zhdanov08} up to astrophysical problems \cite{Ossenkopf91,Perrin90}. In addition, disorder is also used as powerful tool to design materials with tailored optical properties, e.g., exhibiting radiative cooling \cite{Lu17,Yu19,Wong22}. Hence, there is a strong need to efficiently describe light propagation within disordered media. 

Due to the randomness of local arrangements, a description of light transport entirely on a microscopic scale is extremely challenging if not even impossible. However, in many cases using averaged quantities allows for a feasible description of the disordered medium as an effective medium, giving access to macroscopic transport properties. That means a homogenization approach is applied in which the heterogeneous medium with locally varying permittivity is treated as a homogeneous medium possessing an effective permittivity (or equivalently an effective refractive index).

While the effective medium approach enables a relatively simple description of the light transport, the bottleneck of the homogenization is the evaluation of a suitable effective permittivity for arbitrary structures. The first attempts to solve this problem date back more than a century \cite{Brosseau06} and eventually resulted in two famous approaches namely the Maxwell-Garnett (MG) \cite{Garnett04} and the Bruggeman (BG) theory \cite{Bruggeman35}. While both theories only require the permittivity of both constituents and the respective volume fractions, their applicability is restricted to grains much smaller than the incident wavelength due to their quasi-static character \cite{Ruppin00}.

Indeed, the MG theory originally assumes a so-called cermet topology, i.e., separated spheres which are dispersed in a host medium \cite{Garnett04,Markel16,Mallet05}. Thereby, the sphere size parameter $x = ka$ (where $k$ is the wave vector in the host medium and $a$ is the sphere radius) is required to be much smaller than one. In addition, its applicability is generally limited to a small volume fraction of inclusions \cite{Markel16}, although sometimes appropriate results at higher filling fractions can be obtained \cite{Mallet05,Abeles76,Spanoudaki01}.

\begin{figure}[t]
    \centering
    \includegraphics{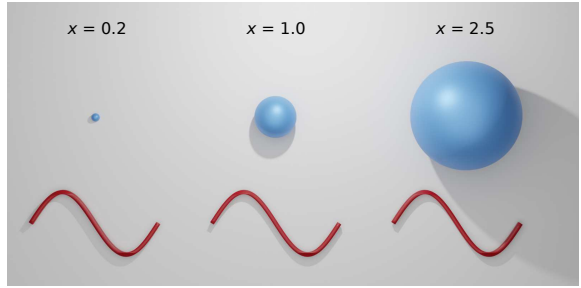}
    \caption{Illustration of different sphere size parameters. The size of a sphere (blue) is displayed in comparison to the wavelength (red) for a typical sphere size parameter in the Maxwell-Garnett regime ($x = 0.2)$, the limit of extended theories ($x = 1.0$) and the largest sphere sizes investigated here ($x = 2.5$). 
    }
\label{fig:sphere_size}
\end{figure}

Nevertheless, due to its simplicity the MG theory is widely used for calculating the effective permittivity for various types of composite media ranging from packings of large spheres \cite{Yazhgur21} even to interconnected structures, e.g., foams, micro-porous media or biological tissues \cite{Cha96,Anguelova08,Wong22,Burresi14,Pompe22}. While the MG theory can provide reasonable results for some structures, its constraints, especially the criterion of small particles, are rarely fulfilled in reality, so its validity is highly questionable in many cases. To circumvent the unpractical restriction of very small spheres several extensions as well as generalized effective medium theories capable to include size effects are presented in the literature \cite{Doyle89,Lakhtakia92,Foss94,Mallet05,Torquato21}. However, these generalized theories are usually limited to a maximum sphere size parameter (or its pendant in the case of non-cermet topologies) of about 0.5 to 1 \cite{Ruppin00,Torquato21}, thus to still somewhat small particles as illustrated in Fig. \ref{fig:sphere_size}. Moreover, it is shown that the most theories work well only in a very limited number of scenarios \cite{Ruppin00,Yu15,Bohren86}.

Here, we present a numerical analysis of the effective refractive index of cermet topologies with sphere sizes up to $x \approx 2.5$ using finite-difference time-domain (FDTD) technique. In contrast to other FDTD approaches \cite{Torquato21,Karkkainen00,Lidorikis07}, we implement the criterion that the forward scattering amplitude of a spherical region of the composite medium vanishes when the homogeneous background possesses a refractive index which matches the effective refractive index of the composite medium \cite{Stroud78,Niklasson81,Chylek83}. 

Using this simple setup, the results of the MG theory are obtained studying a cermet topology with small dielectric particles in a dielectric host. However, increasing the particle diameter the MG theory fails to predict the effective refractive index correctly. Instead, the effective refractive index is found to approximately follow a quadratic polynomial function. Based on the simulation results a simple empirical formula for the coefficients of the polynomial function is derived. Thus, a new mixing rule is reported which allows for calculating the effective refractive index for the yet inaccessible range between $x \approx 1$ and $x \approx 2$.  

\section{Methods}
\subsection{FDTD Simulations}
The software Lumerical FDTD Solutions (Ansys Inc., USA) is used to conduct the FDTD simulations with the setup displayed in Fig. \ref{fig:simulation}(a). Since the software uses a cubic mesh grid as standard, attention must be paid in resolving the spherical shape of the spheres correctly. Therefore, in preliminary tests the mesh size is decreased until the simulation results for the forward scattering amplitude of a single sphere are in agreement with the analytic results of the Mie theory. Thereby, a mesh size of 15\,nm in all directions in combination with the conformal mesh technique is found to be sufficient for all used sphere sizes. The electric field of the injected plane wave has an amplitude of 1\,V/m in all simulations, while the far field is projected onto a hemisphere with a diameter of 1\,m. To determine the effective refractive index of a sphere packing, first of all at least 5 distinct spherical regions are cut out, which possess the same filling fraction than the entire packing. Thereby, the size of the extracted regions is kept constant for different filling fractions, while it is ensured that for the smallest filling fraction 5 spheres are included in the extraction. Subsequently, the extracted regions are placed in the simulation setup and the forward scattering intensity is calculated for different background indices. Since a quadratic dependence between the forward scattering intensity and the background index is found around the point of index matching (see Supplement 1), a parabola is fitted to the simulation results yielding the effective refractive index at its minimum. The final value is obtained by averaging over the distinct extracted regions.

\subsection{Generating Random Sphere Packings}
Random sphere packings are created using a self-written Matlab code (The MathWorks Inc., USA) of the force-biased algorithm  \cite{Moscinski89,Bezrukov02}, with its main idea presented below. After initially placing the origins of $N$ spheres randomly in a container with predefined size, two different diameters are assigned to the spheres. For the inner diameter the maximum value is chosen for which the spheres just do not overlap. The outer diameter is selected such that the sum of all sphere volumes occupies a predefined volume, in most cases the container volume. Thus, overlaps between the spheres are allowed, regarding the outer diameter. Starting with this initial setting the following steps are performed consecutively. First, on every sphere a repulsion force is applied which moves overlapping spheres apart. The calculation of this force is based on a potential function \cite{Moscinski89,Bezrukov02} which relies on the overlap of adjacent spheres according to their outer diameter. In the second step, the outer diameter is gradually decreased. Eventually, the new inner diameter fulfilling the requirement of non-overlapping spheres is computed. Since moving the spheres increases the inner diameter by tendency, the inner and outer diameter approaches. Once the inner diameter reaches the desired value the algorithm is stopped and a random sphere packing of non-overlapping spheres with a specific diameter is obtained. By changing the initial number $N$ of spheres in the container, the filling fraction can be tuned. Here, sphere packings with sphere radii between 100\,nm and 240\,nm are generated, which results in sphere size parameters between $x = 0.90$ and $x = 2.15$ for an incident wavelength of 700\,nm. By varying the wavelength, this range is partially extended up to $x = 2.51$.

\subsection{Parameter of the White Model Structure}
To investigate interconnected structures, a model structure mimicking white \textit{Cyphochilus} scales is used, which is composed of Bragg stacks with a footprint of $300\, \text{nm} \times 300\, \text{nm}$ and a disordered layer thickness as described in ref. \cite{Meiers18}. The center-to-center distance of consecutive dielectric layers is 587\,nm, while the space in-between is vacuum. The variation of the layer thickness is given by a normal distribution which is specified by its mean value $\mu$, its standard deviation $\sigma$ and the interval $I$ of used values. Finally, one third of the layers is randomly omitted, yielding particular bigger gaps between consecutive layers. The effective refractive index of this model structure is calculated analogous to the procedure for the sphere packings. To compute the prediction of the mixing rule for large spheres, the sphere size parameter has to be determined for the model structure. In all cases it is found that taking the sphere size parameter of a sphere, whose volume is equal to the volume of the cuboid with a thickness of $\mu$, delivers excellent agreement between the simulation results and the prediction. 

\section{Results}
\subsection{Validation of the Simulation Setup}

\begin{figure}[p]
    \centering
    \includegraphics{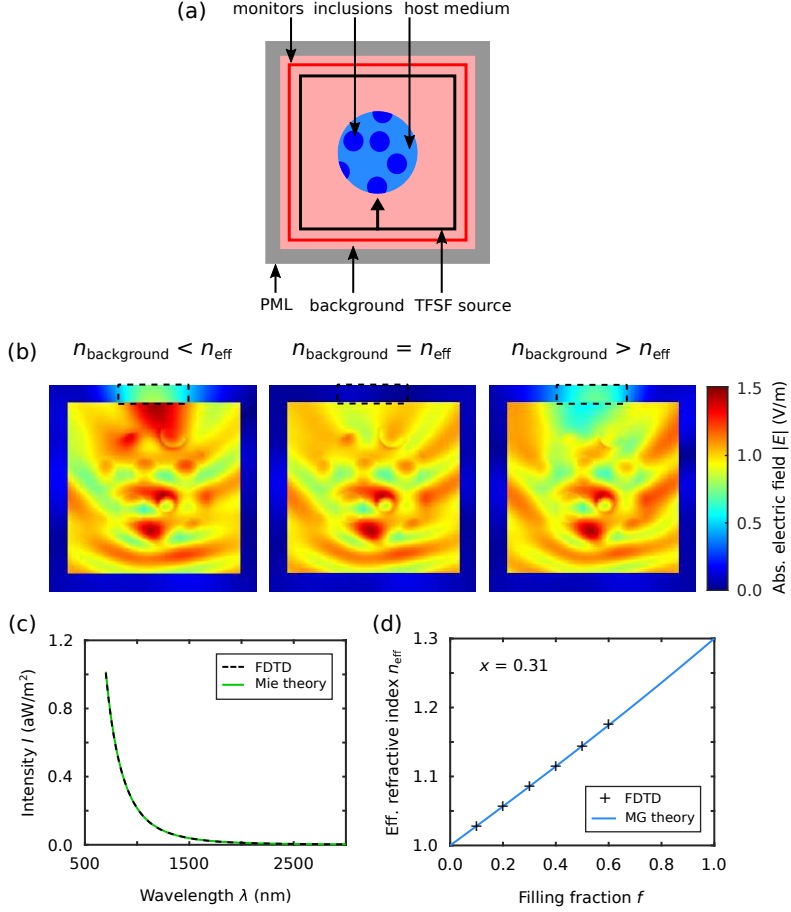}
    \caption{Simulation setup and its validation. (a) 2D cross-section through the 3D simulation region. The simulation region consists of perfectly matched layer (PML) boundaries (gray) limiting the background medium (pale red) in which the total-field scattered-field (TFSF) source (black) is embedded. The composite medium which contains a host medium (pale blue) with spherical inclusions (blue) is illuminated by a plane wave injected at the lower edge of the TFSF source. Outside of the TFSF source, only the scattered portion of light persists, thus the far field projection of the scattered field can be calculated using a box of monitors (red). (b) The distribution of the absolute electric field within the simulation region is shown for three different background indices. The inner square shows the fields inside the TFSF source displaying the impinging plane wave as well as the scattered light, while in the outer rim only the scattered light exists. For a mismatch between the background index and the effective refractive index, forward scattered fields can be seen as marked by the black, dashed box. As shown in Supplement 1 weaker scattering is also observed for other directions. The used composite medium has a filling fraction of 30\% and is composed of spheres with a refractive index of 1.3 and a radius of 180\,nm. (c) Comparison between the far field intensity in the forward scattering direction yielded by FDTD simulation (black) and Mie theory (green) for a single sphere. The sphere has a refractive index of 1.5 and a radius of 100\,nm. (d) Effective refractive index of sphere packings at different filling fractions (black crosses) obtained with the simulation setup shown in (a). The results are in agreement with the prediction of the MG theory (blue curve). For the simulation an incident wavelength of 1000\,nm  as well as spheres with a refractive index of 1.3 and a radius of 100\,nm are used.     
    }
    \label{fig:simulation}
\end{figure}

A 2D cross-section through the applied 3D simulation setup is displayed in Fig. \ref{fig:simulation}(a). The used approach is based on the condition that the forward scattering amplitude vanishes once the refractive index of the background fits the effective refractive index of the composite medium. Therefore, the whole simulation region is filled with a background medium with adjustable refractive index (pale red). In this background a spherical region of the composite medium is embedded, which consists of the host medium (pale blue) and spherical inclusions (blue).

To efficiently determine the forward scattering amplitude the composite medium is placed inside a so-called total-field scattered-field (TFSF) source (black). This source is defined by a 3D box where the total field of the impinging plane wave, i.e., the incident field and the scattered field, is calculated. However, outside of the box only the scattered field exists since the portion of the incident field is subtracted at the edge of the box, as exhibited by the field distribution in Fig. \ref{fig:simulation}(b). Using a box of monitors (red) around the source, the far field projection of the scattered field can be evaluated for all directions. 

In the forward direction the electric field $E_{\text{for}}$ is proportional to the forward scattering amplitude \textit{S}(0°)

\begin{equation}
E_{\text{for}} \sim E_0 \frac{e^{\text{i}kr}}{-\text{i}kr} S(\text{0°}) \,,
\label{equ:E_forward}
\end{equation}

where $r$ is the distance to the scattering particle, $k$ the wave vector and $E_0$ the electric field of the incident wave \cite{Bohren83} (see Methods). In the frequency domain the real field is given by the magnitude of the electric field, thus the intensity $I_{\text{for}}(\lambda) \sim |E_{\text{for}}(\lambda)|^2$ is used for convenience.
The effective refractive index is obtained by varying the background index to minimize the forward scattering amplitude and hence the forward scattering intensity in the far field (see Methods). As displayed in Fig. \ref{fig:simulation}(b), for the case of index matching the suppression of the forward scattering can be observed in the field distribution.

First, the setup is tested by placing a single particle inside the TFSF source and calculating the intensity in the far field for a background index of one. In addition, the scattering amplitude and thus the forward scattering intensity of a single particle is analytically evaluated using Mie theory (e.g. ref. \cite{Bohren83}). Fig. \ref{fig:simulation}(c) shows the comparison between the results obtained by FDTD simulation (black, dashed curve) and Mie theory (green curve) for a sphere with a refractive index of $n_{\text{i}} = 1.5$ and a radius of 100\,nm. It can be seen that for the whole simulated wavelength range both curves are in excellent agreement.

In addition, it also has to be tested if the setup is capable to predict the effective refractive index of a composite medium correctly. Therefore, packings of small spheres with filling fractions between 10\% and 60\% are generated using a slightly modified version of the force-biased sphere packing algorithm \cite{Moscinski89,Bezrukov02} (see Methods). For every filling fraction spherical regions are cut from the entire sphere packing and subsequently used in the simulation setup shown in Fig. \ref{fig:simulation}(a).

Fig. \ref{fig:simulation}(d) shows the simulation results obtained for spheres with $x = 0.314$ and $n_{i} = 1.3$ dispersed in a host medium with $n_{\text{h}} = 1$ at different filling fractions $f$ (black crosses). These findings are compared to the MG mixing rule (blue curve) which reads

\begin{equation}
\epsilon_{\text{MG}} = \epsilon_{\text{h}}\frac{\epsilon_{\text{i}}(1+2f)+2\epsilon_{\text{h}}(1-f)}{\epsilon_{\text{i}}(1-f)+\epsilon_{\text{h}}(2+f)}\,,
\label{equ:MG_mixing_rule}
\end{equation}

with $\epsilon_{\text{i}}$ and $\epsilon_{\text{h}}$ being the permittivity of inclusions and host, respectively \cite{Markel16}. The corresponding effective refractive index is received via $n_{\text{MG}} = \sqrt{\epsilon_{\text{MG}}}$.

As expected for cermet topologies with small spheres, the simulation results are in perfect agreement with the MG theory validating the used simulation geometry.

\subsection{Derivation of a Mixing Rule for Large Particles}
To derive a mixing rule applicable for large spheres, first the size effects are systematically studied by varying the sphere size parameter $x$. Due to $ x = (2\uppi n_{\text{h}}/\lambda)a$, $x$ depends on the sphere radius and the applied wavelength. Hence, both quantities can be used to alter $x$ yielding similar results, see Supplement 1. Here, values between $x = 0.90$ and $x = 2.51$ are realized in random sphere packings (see Methods), which possess filling fractions up to 60\%, close to the limit at which  packings can still be considered random (around 65\% \cite{Moscinski89,Torquato00}).

\begin{figure}[t]
    \centering\includegraphics{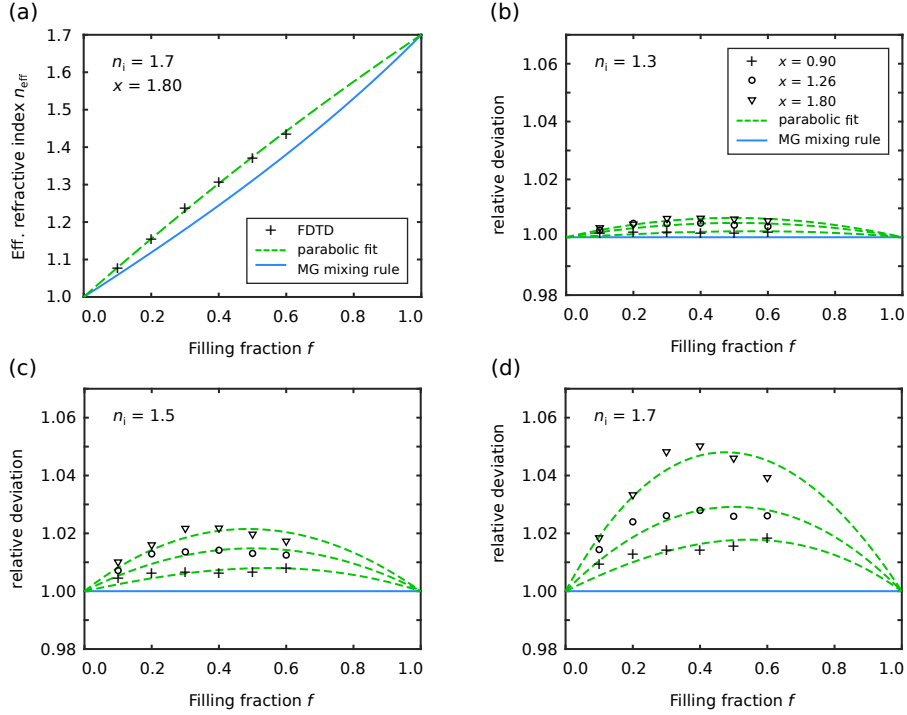}
    \caption{Evaluation of the effective refractive index behavior for different particle sizes. (a) The simulated effective refractive index (black crosses) for a sphere packing with $x = 1.80$, $n_{\text{i}} = 1.7$ and $n_{\text{h}} = 1$ is compared to the prediction of the MG mixing rule (blue curve) and to a parabolic curve (green, dashed), which is fitted to the obtained effective refractive indices. (b)-(d) Deviation of the calculated effective refractive index and corresponding parabolic fits from the MG theory for different sphere sizes and refractive index contrasts of (b) 1.3, (c) 1.5, and (d) 1.7. All quantities are normalized to the results of the MG mixing rule, i.e., the relative deviation from the MG mixing rule is given.
    }
    \label{fig:parabola}
\end{figure}

Fig. \ref{fig:parabola}(a) shows the effective refractive index obtained by FDTD simulations (black crosses) for $x = 1.80$ and a refractive index of $n_{\text{i}} = 1.7$. Here, a significant discrepancy can be seen between the simulation results and the MG mixing rule (blue curve), showing that the latter is not able to correctly predict the effective refractive index for large spheres. Instead, the results can be fitted to a good approximation with a quadratic polynomial function as shown by the green, dashed curve. In general, as sphere size increases, the discrepancy between the effective refractive index and the MG theory increases, with greater refractive index contrast leads to greater discrepancies overall, as shown in Fig. \ref{fig:parabola}(b)-(d).   

In contrast, in all cases the fitted quadratic polynomial describes the obtained behavior closely.
For the polynomial, two assumptions are made considering the limit of very low and high filling fractions. In the limit $f = 0$ the composite medium only consists of the host medium, i.e. it is identical to a homogeneous medium possessing a refractive index of $n_{\text{h}}$. On the other hand, for $f = 1$ the composite medium is solely composed of the inclusion medium with a refractive index of $n_{\text{i}}$.

Based on these findings, the following ansatz for a new mixing rule for large particles is made

\begin{equation}
n_{\text{eff,large}}(f) = p_1f^2 + p_2f + p_3\,,
\label{equ:ansatz}
\end{equation}

where the coefficients $p_1$, $p_2$, and $p_3$ have to be determined. Applying the boundary conditions for $f = 0$ and $f = 1$

\begin{subequations}
\begin{align}
n_{\text{eff,large}}(0) &= p_3 = n_{\text{h}} \,,
\label{equ:low_boundary}
\\
n_{\text{eff,large}}(1) &= p_1 + p_2 + p_3 = n_{\text{i}} \,,
\label{equ:high_boundary}
\end{align}
\end{subequations}

defines two of the three coefficients

\begin{subequations}
\begin{align}
p_2 &= n_{\text{i}} - n_{\text{h}} - p_1\,,
\label{equ:p2}
\\
 p_3 &= n_{\text{h}} \,.
\label{equ:p3}
\end{align}
\end{subequations}

\begin{figure}[t]
    \centering
    \includegraphics{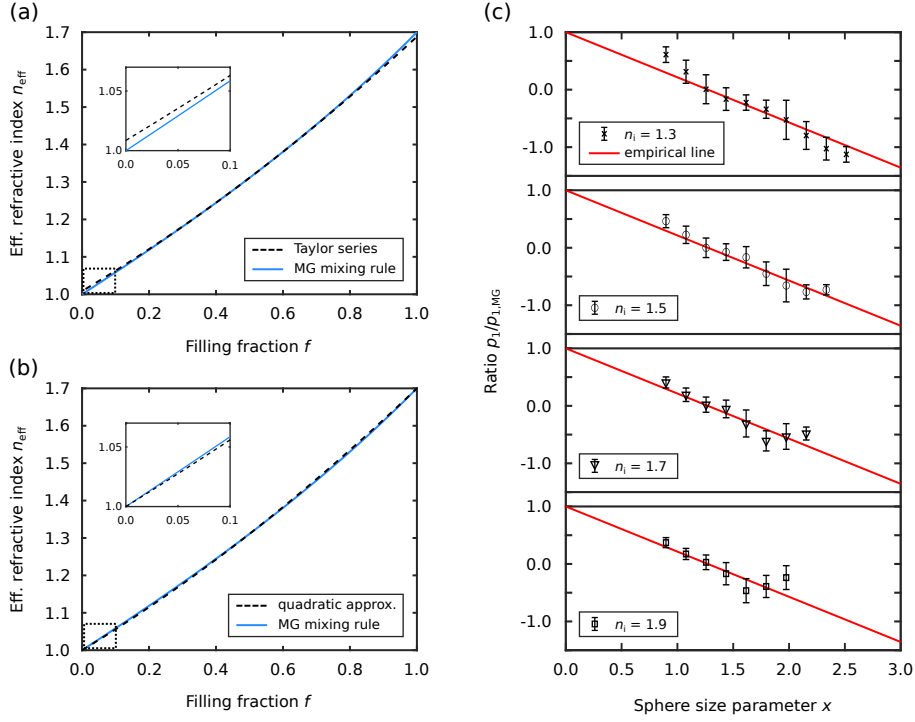}
    \caption{Parameter determination for the mixing rule for large spheres. (a) Comparison between the MG mixing rule for a refractive index of $n_{\text{i}} = 1.7$ (blue curve) and its second order Taylor expansion around $f = 0.5$ (black, dashed curve). The inset shows a close-up of the region marked by the dotted box, revealing that the Taylor series does not fulfill the boundary condition. (b) Corresponding comparison between the MG mixing rule and the quadratic approximation (black, dashed line). The inset displays a close-up of the marked region confirming that the boundary condition is met. (c) Ratio $p_1/p_{\text{1,MG}}$ plotted against the sphere size parameter $x$ for spheres with a refractive index of $n_{\text{i}} = 1.3$, $n_{\text{i}} = 1.5$, $n_{\text{i}} = 1.7$, and $n_{\text{i}} = 1.9$, respectively. The coefficient $p_1$ is obtained via fitting the mixing rule for large spheres to the simulation results while the coefficient $p_{\text{1,MG}}$ is received from the quadratic approximation of the respective MG mixing rule. The error bars display the 95\% confidence interval of the fits. In all cases the host medium is presumed to be vacuum. In addition, the empirically found linear function is displayed (red line), revealing that in the range of interest the ratio follows the same trend independent of the refractive index $n_{\text{i}}$.                
    }
    \label{fig:parameter}
\end{figure}

To find a general expression for $p_1$, the performed quadratic fits are compared to the MG mixing rule. As it can be discerned in Fig. \ref{fig:parabola}(a) the shape of the MG curve resembles a parabola as well. Indeed, performing a second order Taylor expansion of the MG curve around $f = 0.5$ gives a good approximation for almost all filling fractions as shown in Fig. \ref{fig:parameter}(a) (black, dashed curve). However, some deviations are found for very low and high filling fractions, i.e., the boundary conditions mentioned above are not fulfilled (cf. inset). Since these boundary conditions are a basic requirement of the new mixing rule, they should be also enforced for an approximation of the MG mixing rule to allow for comparison. In addition, it is reasonable to demand that the approximation equals the MG mixing rule at $f = 0.5$ as in the case of the Taylor expansion. Hence, this approximation can be given by

\begin{equation}
n_{\text{MG,approx}}(f) = p_{1,\text{MG}}f^2 + p_{2,\text{MG}}f + p_{3,\text{MG}}\,,
\label{equ:parabola_MG}
\end{equation}

with

\begin{subequations}
\begin{align}
&n_{\text{MG,approx}}(0) = p_{3,\text{MG}} = n_{\text{h}} \,,
\label{equ:low_boundary_MG}
\\
&n_{\text{MG,approx}}(0.5) = \frac{1}{4}p_{1,\text{MG}} + \frac{1}{2}p_{2,\text{MG}} + p_{3,\text{MG}} =  n_{\text{MG}}(0.5) \,,
\label{equ:mid_boundary_MG}
\\
&n_{\text{MG,approx}}(1) = p_{1,\text{MG}} + p_{2,\text{MG}} + p_{3,\text{MG}} = n_{\text{i}} \,.
\label{equ:high_boundary_MG}
\end{align}
\end{subequations}

Solving this set of equations yields

\begin{subequations}
\begin{align}
&p_{1,\text{MG}} = 2n_{\text{i}} + 2n_{\text{h}} - 4n_{\text{MG}}(0.5) \,,
\label{equ:p1_MG}
\\
&p_{2,\text{MG}} = -n_{\text{i}} - 3n_{\text{h}} + 4n_{\text{MG}}(0.5) \,,
\label{equ:p2_MG}
\\
&p_{3,\text{MG}} = n_{\text{h}} \,.
\label{equ:p3_MG}
\end{align}
\end{subequations}

Using Eq. (\ref{equ:parabola_MG}) with the coefficients (\ref{equ:p1_MG}) to (\ref{equ:p3_MG}) indeed closely approximates the MG mixing rule while fulfilling the boundary conditions (see Fig. \ref{fig:parameter}(b)).

As it is exhibited in Fig. \ref{fig:parabola} the coefficient $p_1$ depends on $n_{\text{i}}$ and $x$. However, within the scope of the mixing rule for large spheres (between $x \approx 1$ and $x \approx 2$) the explicit dependence on $n_{\text{i}}$ can be overcome by considering the ratio $p_1/p_{1,\text{MG}}$ instead of $p_1$. As shown in Fig. \ref{fig:parameter}(c) the ratio $p_1/p_{1,\text{MG}}$ indeed follows the same trend (red line) for different $n_{\text{i}}$, i.e., the $n_{\text{i}}$ dependence of the mixing rule for large spheres can be ascribe to the one of the MG mixing rule. Within the scope the ratio $p_1/p_{1,\text{MG}}$ is empirically found to be related to the sphere size parameter by the linear function

\begin{equation}
\frac{p_1}{p_{1,\text{MG}}} = 1 - \frac{\uppi}{4}x\,,
\label{equ:empirical}
\end{equation}

as indicated by the red line in Fig. \ref{fig:parameter}(c). Inserting Eq. (\ref{equ:p1_MG}) in Eq. (\ref{equ:empirical}) $p_1$ can be computed. Thus, within the scope $x \approx 1$ to $x \approx 2$ the mixing rule for large sphere is generally give by

\begin{equation}
n_{\text{eff,large}}(f) = p_1f^2 + (n_{\text{i}} - n_{\text{h}} - p_1)f + n_{\text{h}}\,,
\label{equ:new_mixing_rule}
\end{equation}

with

\begin{equation}
p_1 = (1 - \frac{\uppi}{4}x)\cdot(2n_{\text{i}} + 2n_{\text{h}} - 4n_{\text{MG}}(0.5))\,.
\label{equ:p1_final}
\end{equation}

While so far the host medium is assumed to be vacuum, the empirical formula (\ref{equ:empirical}) holds also true for host media with $n_{\text{h}} \neq 1$ (see Supplement 1) validating the mixing rule for large spheres in its general form as given above.

\subsection{Comparison of Different Mixing Rules}
\begin{figure}[h]
    \centering
    \includegraphics{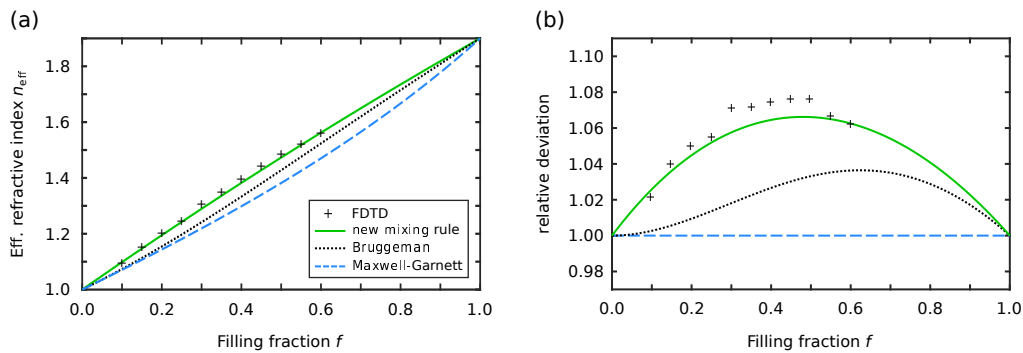}
    \caption{Comparison of different mixing rules. (a) The prediction of the mixing rule for large spheres (green, solid curve) is compared to the respective results of the Bruggeman (black, dotted curve) and the Maxwell-Garnett mixing rule (blue, dashed curve) for a sphere radius of 190\,nm and a refractive index of $n_{\text{i}} = 1.9$. The host medium is assumed to be vacuum and the incident wavelength is set to 700\,nm yielding a sphere size parameter of $x = 1.71$. (b) Relative deviation of the quantities shown in (a) from the prediction of the MG mixing rule.      
    }
    \label{fig:comparison}
\end{figure}

The prediction (calculated with Eq. (\ref{equ:new_mixing_rule})) of the here derived mixing rule for large spheres is shown in comparison to the respective results of the MG and the Bruggeman mixing rule as displayed in Fig. \ref{fig:comparison}. For the used index contrast of 1.9 and sphere size $x = 1.71$, it can be seen that only the mixing rule for large spheres describes the obtained effective refractive index closely over a great range of filling fractions. In contrast, the MG theory clearly underestimates the effective refractive index, revealing a deviation of around $\Updelta n = 0.1$ for large filling fractions between 30\% and 60\%. In this filling fraction range the prediction of the BG theory is better than that of the MG theory, although there is still an underestimation of about $\Updelta n = 0.05$.

The difference between the MG and the BG theory can be understood, considering their original purposes. The MG theory is developed for separated grains which are dispersed in a host medium, hence only weak interaction between grains is assumed. Contrarily, the BG theory presumes an aggregate structure, i.e., space which is randomly occupied by two different materials. Thereby, every constituent is modeled as a tight arrangement of homogeneous, small spheres \cite{Chylek83}. Although so far only separated grains in a host medium are considered, the character of these structures for large filling fractions resembles rather an aggregate structure. This can be unambiguously seen for filling fractions above 50\% where the volume of the `host' medium is lower than the volume of the `inclusions', rendering this assignment unsuitable. The approximation as an aggregate structure holds also true for filling fractions somewhat below 50\%, where many spheres are in close proximity, i.e., forming clusters. In consequence, for large filling fractions the BG outperforms the MG mixing rule; however, both theories are unable to correctly predict the effective refractive index, which is only achieved by the mixing rule for large spheres. 

While large filling fractions lead to aggregate-like structures, the spheres are still generally non-interconnected as well as monodisperse in size. In contrast, many composite media, e.g. biological tissue, consist of disordered, interconnected structures with varying feature sizes \cite{Schmitt96}. Therefore, the predictions of different mixing rules are tested in the context of more realistic, interconnected structures.

One example of such a structure are the white scales of beetle \textit{Cyphochilus}. Inside these scales a disordered network composed of chitin struts with various thickness is found. As described in ref. \cite{Meiers18} it is possible to model this network by replacing the struts with simple blocks of fitting thickness. This yields a simple model structure which consists of partly interconnected cuboids with a normally distributed thickness (see Methods) and therefore is in distinct contrast to topologies considered so far. Using the FDTD setup, the effective refractive index of this model structure is calculated and compared to the results of different mixing rules.

\renewcommand{\arraystretch}{1.5}
\begin{table}[t]
\centering
\caption{Comparison between the effective refractive index of a model structure for white beetle scales and the prediction of different mixing rules. The normal distribution of the structure's layer thickness is given by the mean value $\mu$, the standard deviation $\sigma$ and the interval $I$ of the used thicknesses. A refractive index of $n_{\text{i}}$ is applied and the resulting filling fraction is denoted by $f$. The effective refractive index $n_{\text{FDTD}}$ obtained by FDTD simulation is compared to the predictions of the mixing rule for large spheres $n_{\text{eff,large}}$, the BG mixing rule $n_{\text{BG}}$ as well as to the MG mixing rule $n_{\text{MG}}$.\\\label{tab:comparison}}

\begin{tabular}{ |c|c|c|c|c|c|c|c|c| }
\hline
$n_{\text{i}}$ & $\mu$ & $\sigma$ & $I$ & $f$ & $n_{\text{FDTD}}$ & $n_{\text{eff,large}}$ & $n_{\text{BG}}$ & $n_{\text{MG}}$ \ \\
\hhline{|=|=|=|=|=|=|=|=|=|}
1.5 & 190 nm & 80 nm & [50 nm, 330 nm] & 25\% & 1.126 & 1.127 & 1.116 & 1.113\\
\hline
1.5 & 230 nm & 160 nm & [50 nm, 410 nm] & 30\% & 1.151 & 1.153 & 1.141 & 1.136\\
\hline
1.5 & 370 nm & 160 nm & [190 nm, 550 nm] & 40\% & 1.210 & 1.207 & 1.191 & 1.183\\ 
\hline
1.7 & 130 nm & 80 nm & [50 nm, 210 nm] & 15\% & 1.115 & 1.105 & 1.092 & 1.088\\ 
\hline
1.7 & 190 nm & 80 nm & [50 nm, 330 nm] & 25\% & 1.179 & 1.179 & 1.158 & 1.149\\ 
\hline
\end{tabular}
\end{table}

As displayed in Table \ref{tab:comparison} for different model parameters and refractive indices, the mixing rule for large spheres predicts the effective refractive index closely in all cases. In contrast, the results of the MG and BG mixing rule reveal notable deviations with smaller discrepancy in the case of the BG mixing rule, as expected for aggregated structures. Overall, this clearly indicates that the new mixing rule is not only capable to include size effects correctly but is also applicable for various structure types, ranging from separated grain structures to more realistic aggregated structures.

\section{Discussion}
As shown in the previous section, the here presented mixing rule for large spheres can be successfully deployed in a large number of scenarios. Nevertheless, there are still some limitations of its applicability, given by the scope of the empirical formula. As exhibited in Supplement 1, the empirical formula rather depends on the refractive index contrast than on the actual value of $n_{\text{i}}$ and $n_{\text{h}}$, which confines the discussion to the influence of the index contrasts on the scope.

For $x \rightarrow 0$ the mixing rule for large spheres transitions into the quadratic approximation of the MG mixing rule as expected for vanishing sphere size. However, as long as the dipole polarization of the inclusions can be given by its electrostatic value, the MG theory delivers the correct effective refractive index \cite{Ruppin00,Markel16}.
Thus, also for $x > 0$ the MG mixing rule can yield correct results (cf. Fig. \ref{fig:simulation}(d) for $x \approx 0.3$). In consequence, for sphere size parameters near the electrostatic regime the ratio $p_1/p_{\text{1,MG}}$ is expected to be closer to one than predicted by the empirical formula, limiting the scope at the lower edge. This behavior is more pronounced for small index contrasts (cf. Fig. \ref{fig:parameter}(c)), where the electrostatic description holds true for larger sphere sizes. 

At the upper edge of the scope, it can be observed that at a certain point the ratio starts to stay constant or even increases and hence deviates from the empirical formula. This point is found to shift towards smaller sphere size parameter when the index contrast is increased, which limits the scope to $x~\lesssim~2$ for index contrasts up to $n_{\text{i}}/n_{\text{h}} = 2$. Even larger index contrasts diminish the scope further as exemplarily shown in Supplement 1 for $n_{\text{i}}/n_{\text{h}} = 2.8$.

The upper limit of the scope can be understood considering different realizations of the same sphere packing. As shown in Supplement 1, for small spheres the obtained effective refractive index barely varies for different realizations. In contrast, for sphere size parameters above the scope the gained values spread noticeably. This indicates that the individual arrangement of spheres becomes dominant for the scattering behavior of very large spheres. Consequently, the description as an effective medium becomes invalid, which limits the applicability of any mixing rule.

The exact determination of the effective refractive index is crucial in many application. Since cancerous and healthy tissue possess different refractive indices \cite{Wang11,Yamaguchi16,Cassar21}, it could be recently shown that modeling biological tissue as an effective medium allows to determine between malign and healthy tissue at an early stage of the cancerous disease \cite{Gric20,Gric22}. To delimit the tumor, precise determination of the effective permittivity is needed, which requires more accurate mixing rules than the currently used MG mixing rule \cite{Gric22}, especially when size effects are not negligible. This is the case for the frequently used THz regime, where the typical size of human cells (10 - 100\,µm \cite{Campbell06}) can be in the range of used wavelengths (about 3\,mm - 40\,µm for 0.1 - 7\,THz) while index contrasts between malign and healthy tissue are up to 1.8 at these frequencies \cite{Cassar21}.       

Another example where  size effects are not negligible are white paint formulations, which possess typical refractive index contrasts and sphere size parameters, both in the order of 1.8 \cite{Pattelli18}. Hence, in the broad field of scattering and whiteness optimization \cite{Pompe22,Pattelli18,Jacucci19,Jacucci21}, incorporating size effects in the calculation of the effective refractive index is of high interest.

\section{Conclusions}
A new mixing rule for calculating the effective refractive index of composite media is presented, which is based on a quadratic function whose coefficients can be computed using an empirical formula. Although being of a rather simple form, the new mixing rule provides correct results for large particles in the range of $x \approx 1$ to $x \approx 2$, expanding the current upper limit for such rules ($x \lesssim 1$) to sphere sizes close to the regime of geometrical optics. While the mixing rule is derived for cermet topologies, testing for more complex structures yields very accurate predictions which clearly outperform the results of common mixing rules. The new mixing rule is therefore believed to supersede the MG and BG mixing rule in numerous situations since natural as well as many artificial disordered media rather fall in the scope of the here presented rule than in the scope of the other mixing rules. So far only non-absorbing, dielectric materials are studied which occur in many applications. Nevertheless, the presented simulation setup can be extended to investigate also absorbing and metallic particles, which will be conducted in further studies.

\pagebreak

\textbf{Funding}

This work was founded by Deutsche Forschungsgemeinschaft (DFG), project no. 255652081.

\bigskip
\textbf{Acknowledgments}

The authors gratefully acknowledge financial support from the German Research Foundation DFG within the priority program "Tailored Disorder -  A science- and engineering-based approach to materials design for advanced photonic applications" (SPP 1839). 

\bigskip
\textbf{Disclosures}

The authors declare no conflicts of interest.

\bigskip
\textbf{Data Availability Statement}

All data are available in the main text or the supplemental document. Additional data related to this paper may be obtained from the authors upon reasonable request.

\bigskip
\textbf{Supplemental document}

See Supplement 1 for supporting content.




\newpage

\renewcommand{\thesection}{S\arabic{section}}
\renewcommand{\thefigure}{S\arabic{figure}}
\setlength{\parindent}{0em}

\onehalfspacing

\begin{center}
    {\Large \textbf{Mixing Rule for Calculating the Effective Refractive Index Beyond the Limit of Small Particles}}
\end{center}

\medskip

\begin{center}    
    {\Large Supplemental Document}
\end{center}

\medskip

\begin{center}
    {\large D. T. Meiers and G. von Freymann}
\end{center}

\bigskip
\bigskip
\bigskip
\bigskip
\bigskip

\begin{figure}[hb]
    \centering
    \includegraphics{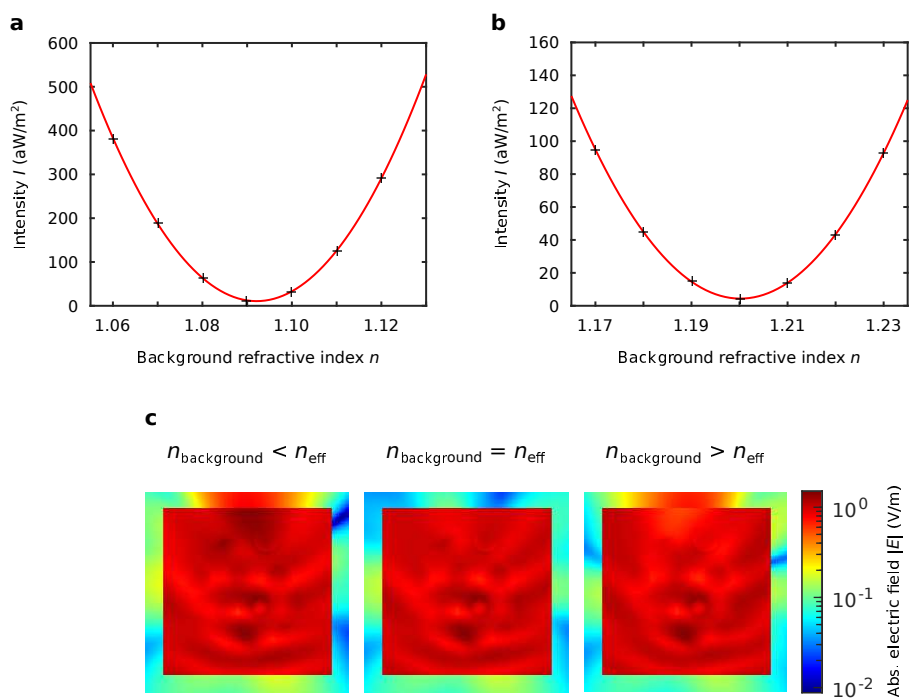}
    \caption{Determination of the effective refractive index based on the simulation results. a) Forward scattering intensity of a spherical region extracted from a sphere packing composed of spheres with a radius of 180\,nm, a refractive index of 1.3, and a filling fraction of 30\%, for different background indices (black crosses). The results can be fitted with a parabola (red curve) in excellent agreement. The minimum of the parabola marks the refractive index match between the composite medium and the background, hence yielding the effective refractive index. The minimum forward scattering intensity is slightly greater than zero due to unavoidable numerical errors. b) Same as (a) but for a sphere packing with a filling fraction of 40\% composed of spheres with a radius of 140\,nm and a refractive index of 1.5. c) Logarithmic plot of the corresponding field distributions for the sphere packing considered in (a) and three different background indices. For index matching between the background index and the effective refractive index no predominant scattering direction can be identified, while pronounced forward scattering can be observed in the case of mismatch.
    }
    \label{fig:determination_n_eff}
\end{figure}

\begin{figure}[ht]
    \centering
    \includegraphics{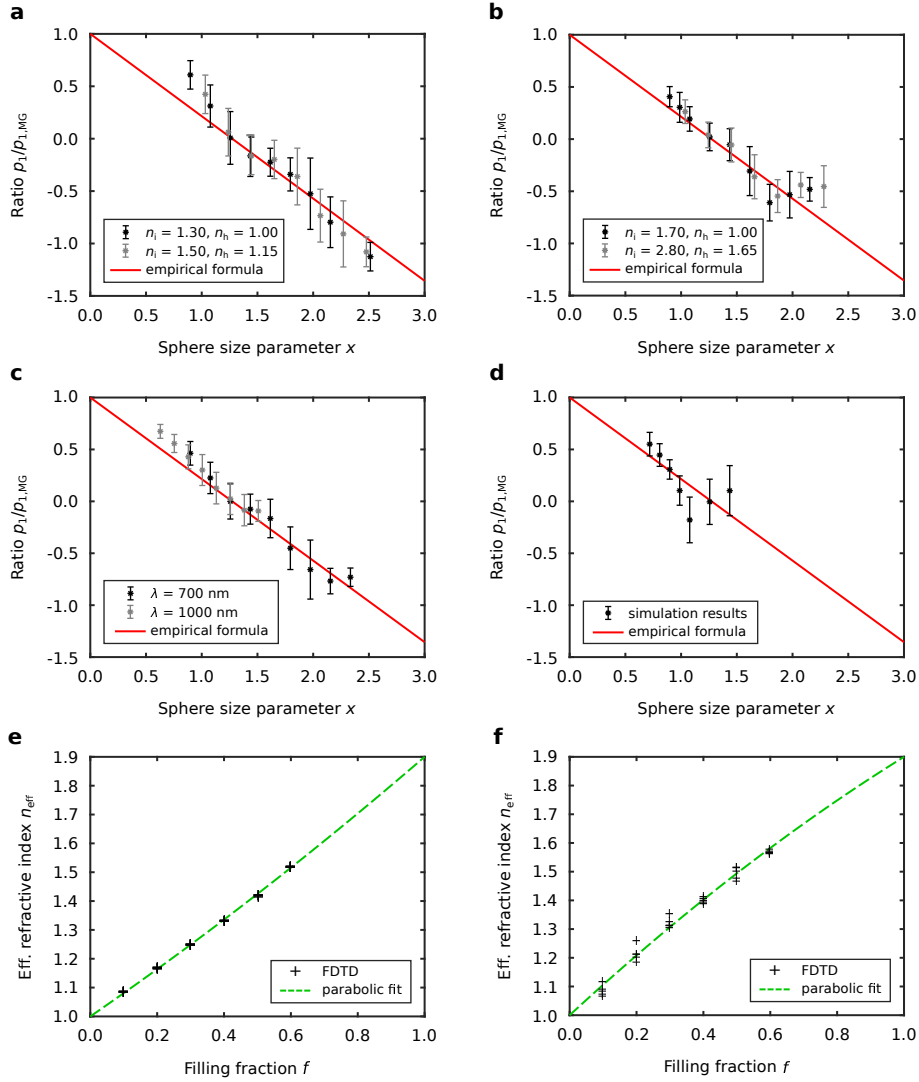}
    \caption{Applicability and limitations of the empirical formula for different refractive index contrasts and wavelengths. a) Comparison between the empirical formula and the ratio $p_{1}/p_{\text{1,MG}}$ for two different $n_{\text {i}}$ and $n_{\text {h}}$ configurations, which exhibit a similar index contrast of $n_{\text {i}}/n_{\text {h}} \approx 1.3$. Independent of the used configuration the same trend is obtained for comparable index contrast. b) Same as (a) but for two configurations possessing a contrast of $n_{\text {i}}/n_{\text {h}} \approx 1.7$. c) Obtained ratio $p_{1}/p_{\text{1,MG}}$ for a wavelength of 700\,nm and 1000\,nm, respectively, of the incident light. In both cases the ratio follows the same trend, revealing that the ratio only depends on the sphere size parameter and not on the explicit choice of wavelength and radius. d) Ratio $p_{1}/p_{\text{1,MG}}$ plotted against the sphere size parameter for a large refractive index contrast of $n_{\text {i}}/n_{\text {h}} = 2.8$. The upper edge of the scope lies around $x \approx 1$, showing the limitation of the description as an effective medium in the case of large index contrasts. e) Effective refractive index obtained for 5 different extracted regions of the same sphere packing with a sphere size of 100\,nm as well as a refractive index of $n_{\text {i}} = 1.9$. Different extracted regions and hence different local arrangements exhibit only marginal variation of the effective refractive index, as also depicted in comparison with the parabolic fit. f) Same as (e) but for spheres with a large radius of 240\,nm, which is beyond the identified scope. Distinct extracted regions display notable differences in the effective refractive index, revealing that the assumption of an effective refractive medium becomes invalid.  
    }
    \label{fig:applicability_empirical_formula}
\end{figure}

\end{document}